\newcommand{\be}{\begin{equation}}
\newcommand{\ee}{\end{equation}}
\newcommand{\ba}{\begin{eqnarray}}
\newcommand{\ea}{\end{eqnarray}}
\newcommand{\non}{\nonumber\\ }
\newcommand{\eq}[1]{Eq.(\ref{#1})}
\newcommand{\Hil}[1]{{(C^2)^{[#1]}}}
\newcommand{\vE}{{\vec{\cal E}}}

\documentstyle[12pt,fullpage]{article}
\input epsf
\begin{document}

\renewcommand{\thefootnote}{\fnsymbol{footnote}}
\font\csc=cmcsc10 scaled\magstep1
{\baselineskip=14pt
 \rightline{
 \vbox{\hbox{UT-821}
       \hbox{July 1998}
}}}

\vfill
\begin{center}
{\large\bf
Matrix Theory, Hilbert Scheme
and Integrable System
}

\vfill
{\csc Yutaka MATSUO}\footnote{
      e-mail address : matsuo@phys.s.u-tokyo.ac.jp},
\vskip.1in

{\baselineskip=15pt
\vskip.1in
 Department of Physics, University of Tokyo\\
 Hongo 7-3-1,  Bunkyo-ku, Tokyo, Japan
\vskip.1in
}

\end{center}
\vfill

\begin{abstract}
{
We give a reinterpretation of the matrix
theory discussed by Moore, Nekrasov and Shatashivili (MNS)
in terms of the second quantized operators
which describes the homology class of
the Hilbert scheme of points on surfaces.
It naturally relate the contribution from each pole
to the inner product of orthogonal basis of
free boson Fock space. 
These basis can be related to the eigenfunctions
of Calogero-Sutherland (CS) equation
and the deformation parameter of MNS is identified with
coupling of CS system.
We discuss the structure of Virasoro symmetry in this model.
}
\end{abstract}
\vfill

hep-th/9807085
\setcounter{footnote}{0}
\renewcommand{\thefootnote}{\arabic{footnote}}
\newpage
\vfill

%
\section{Introduction}

It is widely believed that the
method of the second quantized oscillators
captures the essence of the dynamics of
$p$-branes. This method was based on 
the observation that the homology group structure
of the Hilbert scheme of points on surfaces
can be essentially given by free fields oscillators 
\cite{r:Nakajima}.  
One of the  most cerebrated examples
is the  calculation of
black hole entropy \cite{r:SV} by using 
Vafa-Witten formula \cite{r:VF}.
It was also discussed in the connection with
the matrix string theory \cite{r:DVV}, 
GKM algebra\cite{r:HM}, five-brane quantization
\cite{r:Nakatsu}, the generalization to elliptic
genus \cite{r:DMVV}, and more recently, in relation with
AdS/CFT correspondence, \cite{r:V}\cite{r:deB}.
Whereas the matrix models \cite{r:BFSS}\cite{r:IKKT}
essentially give the first quantized description of
$p$-branes, this method gives the second quantization
of the branes since the free field oscillators are
supposed to create/annihilate the (degenerate) $p$-branes.

In this paper, we point out that
this technique may be also useful to explore the relation
between the matrix model and the integrable systems.
We will use the result of 
Moore, Nekrasov and Shatashivili \cite{r:MNS}
who argued that the D-instanton matrix integration
can be regarded as cohomological field theory
of the Hilbert scheme of points on surfaces.
By using this relation they
performed the matrix integration explicitly and
found that the partition function is given as
a sum over contributions from each homology class. 
They considered three versions of the matrix integrations,
reductions from 4, 6, 10 dimensional super 
Yang-Mills theories.
The second one is the example directly related to
Nakajima's construction \cite{r:Nakajima}.
Contribution from each homology class is
given as fixed points of the $U(1)$-transformations
and is labeled by Young tableau.

We find that each contribution from Young tableau
is directly related to the correlation function
of free fermions associated with
Nakajima's free bosons
when we choose appropriate
value for MNS's ``deformation parameters''.
When these parameters are general,
we need to replace free fermionic state to
Jack polynomial \cite{r:St}\cite{r:Mac}.
They are originally introduced to represent
eigenfunctions of Calogero-Sutherland model\cite{r:CS}
and their bosonized form was studied in \cite{r:AMOS}.
From representation theory viewpoint, they are associated
with the singular vector of ${\cal W}_N$ algebras
with central charge $c<N$ and gives a deformation
of the basis of Hilbert space of free bosons.

Finally we make a comparison of oscillator representation
of new matrix theory with the matrix models \cite{r:old}
associated with $c\leq 1$ quantum gravity.
In this model, the underlying structure of free
field appears in the form of Dyson-Schwinger
equation \cite{r:FKN},
$
L_n Z=0.
$
Similar relation was also found in quantum cohomology
\cite{r:QC} where free fields are associated with
the cohomology of target space.
In this respect, the quantum cohomology and 
Hilbert scheme have precisely similar structure\cite{r:Nakajima}.
We hope that our analysis may clarify some aspects of
this correspondence in the future.

\section{Hilbert scheme of points on surfaces}
Let us give a brief review of free oscillator
method for the Hilbert scheme of points on surfaces\cite{r:Nakajima}.

Consider a typical equation of motion for 
the matrix model,
\begin{equation}
 \left[B_1,B_2\right]=0,
\end{equation}
where $B_i\in GL(N,C)$ ($i=1,2$) with gauge symmetry 
$B_i\rightarrow gB_ig^{-1}$.
It may be regarded as describing ``motion'' of $N$ D-instantons
in $C^2$ whose locations are described as the eigenvalue of $B_i$.
Something very interesting happens when some of D-instantons
collide.  In such a situation, we can not simultaneously
diagonalize these two matrices.  As an example, let us
consider $n$ D-instantons sitting at the origin.
One may take the above matrices as,
\begin{equation}
\label{e_B}
(B_1)_{ij}=\delta_{i,j+1}\quad (j=1,\cdots,N-1),
\quad
B_2=\sum_{j=1}^{N-1} a_j B_1^{j}.
\end{equation}
The off diagonal components $a_1,\cdots,a_{N-1}\in C$ 
describes the mutual ``angle'' of $N$ points
which are located infinitesimally close.
They gives rise to
$2(n-1)$-cycles
generated by the blowup of the orbifold singularity of the 
symmetric product $(C^{[2]})^N/S_N$.
Such an object is called
as the Hilbert scheme of points on $C^2$
and denoted as $\Hil{N}$. 

In the beautiful
lecture note, Nakajima described that the Homology group
can be organized by the free boson oscillators
if we introduce ``the generating space'' 
$\oplus_{N=1}^\infty\Hil{N} $.
Such a construction is motivated by the fact that
the generating functional of Poincare polynomial is given by
\begin{eqnarray}
 \sum_{n=0}^\infty q^n P_t(\Hil{n})& =& \prod_{n=1}^\infty
  \frac{1}{(1-t^{2n-2}q^n)}\nonumber\\
&=& 1+q+(t^2+1)q^2+(t^4+t^2+1)q^3+\cdots.
\end{eqnarray}
Intuitively, $n$-th operator $P_{-n}$ ($n>0$) describes
insertion of $n$ coincident D-instantons attached with
$2n-2$ cycle described above.

To calculate the homology group, one may use the
$T^2$ action on $\Hil{N}$,
$
(B_1,B_2)\rightarrow (z_1B_1, z_2B_2),
$
and use the fixed point theorem.
The fixed points are defined by,
\begin{equation}
z_1 B_i= gB_i g^{-1}\quad i=1,2
\end{equation}
for certain $g$.
Such a set of $B_i$ give a decomposition
of vector space $C^N$.
The fixed points are labeled by
Young diagram of $N$ boxes where each box describes
an eigenstate with respect to the $T^2$ action.
$B_1$ map each box to the box below and $B_2$ to the
right. The action of $B_1$ (resp $B_2$) can be alternatively 
interpreted as deleting the first row (column) of the Young diagram.


In analogy with the fermionic construction of
Grassmanian homology group, we propose to associate
each fixed point with the fermion Hilbert space by using
Maya diagram\cite{r:Sato}.  We give an illustration of
such mapping in Figure 1.
In terms of fermions, the action of $B_1$ (resp $B_2$) is 
equivalent to removing the top excited state 
(resp. filling the lowest unfilled state)\footnote{
We have to mention that,
in \cite{r:Nakajima}\cite{r:Nakajima2}\cite{r:Lehn},
the vertex operator of free boson oscillator is introduced
in terms of curves embedded in the target space.
}.

\vskip 4mm  
\centerline{\epsfbox{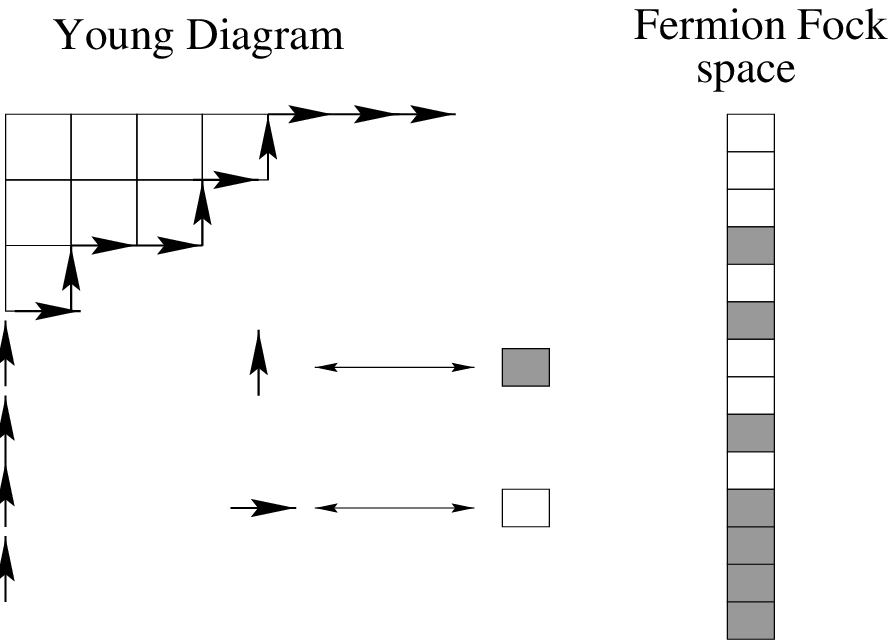}}
\centerline{{\bf Figure 1:} Maya diagram}

\section{Topological Matrix Integration}

In general situation, the second
quantized operator describes only the topological sector
of the matrix theory. 
Recently, in connection with computing
the Witten index of BFSS matrix model, Moore, Nekrasov and
Shatashivili \cite{r:MNS}
have shown that the matrix integrations
associated with dimensional reduction of 10, 6, and 4 dimensional
super Yang-Mills theory
are essentially topological,
i.e. the action is given by the BRST boundary.
The BRST transformation can be typically written
as, $ QB_i = \psi_i$, $ Q\psi_i = [\phi,B_i]$.
We  denote $\phi=X_{D-1}+i X_D$ (longitudinal components)
and $B_j=X_{2j-1}+iX_{2j}$ ($j=1,\ldots,D/2-1$)
(transversal components).  They also rearrange fermions as
$\Psi\rightarrow \Psi_a=(\psi_j, \psi^\dagger_j),\vec\chi, \eta$.
The bosonic part of equation of motion for transversal 
degree of freedom is
\begin{eqnarray}
 &D=4 & \quad \vE = [B_1,B_1^\dagger]\nonumber\\
 &D=6& \quad \vE=\left([B_1,B_1^\dagger]+[B_2,B_2^\dagger],
[B_1,B_2], [B_2^\dagger,B_1^\dagger]\right)\\
 &D=10& \quad \vE=\left([B_i.B_j]+\frac{1}{2}
\epsilon_{ijkl}[B_k^\dagger.B_l^\dagger], i<j, \sum_i
[B_i,B_i^\dagger]\right)
\end{eqnarray}
They represent respectively complex ($D=4$), quarternion ($D=6$),
and octonionic ($D=10$) structure of the transversal space.
$D=6$ case is directly related to discussion 
in the previous section.

In these cases, there is a hope that Nakajima's
second quantized operator essentially describe
the physical degree of freedom of matrix theory.
MNS carried out the major part of matrix integration
except for diagonal part of $\phi$. 
\begin{eqnarray}
 I_{D=4} & = & \frac{N}{N!E_1^{N-1}}\int\prod_i d\phi_i
  \prod_{i\neq j}\frac{\phi_{ij}}{(\phi_{ij}+E_1+i0)}\nonumber\\
 I_{D=6}& = & \left(\frac{E_1+E_2}{E_1E_2}\right)^{N-1}
  \frac{N}{N!}\int \prod_i d\phi_i
  \prod_{i\neq j}\frac{\phi_{ij}(\phi_{ij}+E_1+E_2)}{
  \prod_{\alpha=1,2}(\phi_{ij}+E_\alpha+i 0)}\\
 I_{D=10} & = & \left(\frac{(E_1+E_2)(E_2+E_3)(E_3+E_1)}{
		 E_1E_2E_3E_4}\right)^{N-1}
  \int\prod_i d\phi_i \prod_{i\neq j}\frac{P(\phi_{ij})}{Q(\phi_{ij})}
  \nonumber\\
 P(x)& = & x(x+E_1+E_2)(x+E_2+E_3)(x+E_1+E_3)\nonumber\\
 Q(x) & = & \prod_{\alpha=1}^4(x+E_\alpha+i0)\nonumber
\end{eqnarray}
where $E_\alpha$ ($\alpha=1,2$ for $D=6$ and $1,2,3,4$ for $D=10$ case)
are ``deformation'' parameters and $\phi_{ij}=\phi_i-\phi_j$.
In $D=10$ case, they are constrained by
$\sum_\alpha E_\alpha=0$. These parameters are introduced to 
reguralize the behavior of integral when all branes are collapsed a
single point $(\phi_{ij}=0)$. 
The pole of the integrand is then given by the condition 
$\phi_{ij}=E_\alpha$ and all of them are.

\subsection{$D=4$ case}
The location of pole is described as,
\begin{equation}
 \phi_{\sigma_{i+1}}=\phi_{\sigma_i}+E_1+0i,
\end{equation}
with appropriate permutation $\sigma$.
This is the solution to the equation,
$[B,\phi]=E_1 B$ if we identify $B=B_1$ in \eq{e_B}.
Namely $B$ is the fixed point.

\subsection{$D=6$ case}
We have two separation parameters $E_\alpha$ 
($\alpha=1,2$)
and the pole is conveniently parameterized by Young tableau,
$N=\nu_1+\cdots+\nu_{\nu'_1}=\nu'_1+\cdots+\nu'_{\nu_1}$.
Let $(\alpha,\beta)$ denote the position of a box in
Young tableau, $1\leq\alpha\leq \nu_\beta$,
$1\leq\beta\leq \nu'_\alpha$. Then the pole location is
parameterized by
\begin{equation}
 \phi_{(\alpha,\beta)}=(\alpha-1)E_1+(\beta-1)E_2.
\end{equation} 

MNS evaluated the contribution of poles parameterized by
Young tableau $D$,
\begin{eqnarray}\label{eqSUM}
&&Y_D=(-1)^{N-1}E_1 E_2\label{eqYD}\\
 &&\cdot\frac{
 \prod_{\alpha,\beta\neq (1,1)} ((\alpha-1)E_1+(\beta-1)E_2)(\alpha E_1
 +\beta E_2)}{\prod_{(\alpha,\beta)}((\nu_\beta
 -\alpha+1)E_1+(\beta-\nu'_\alpha)E_2)((\alpha-\nu_\beta)E_1
 +(\nu'_\alpha-\beta+1)E_2)}.\nonumber
\end{eqnarray}
Since the integration should not 
depend on the deformation parameters,
the sum of $Y_D$ is independent of $E_\alpha$,
\begin{equation}
 \frac{1}{N}\sum_{|D|=N}Y_D=\frac{1}{N^2}.
\end{equation}

In view of complexity of \eq{eqYD}, this result seems highly 
non-trivial. However, as we can immediately notice, the factor
in the denominator is the factor which appear in the normalization
of Jack polynomial \cite{r:St}.  We will discuss this issue in
the next section.

\subsection{$D=10$ case}
The Young tableau is generalized to
``3-dimensional partition'',
\begin{equation}
 N=\sum_{i_1,i_2,i_3\geq 0}\nu_{i_1,i_2,i_3}
\quad \mbox{where}\quad
n_{i_1,i_2,i_3}\geq n_{j_1,j_2,j_3}
\end{equation}
whenever $i_\alpha\leq j_\alpha$.
In the language of higher dimensional partition,
the poles of $D=4,6,10$ integral is parameterized respectively by
$k=0,1,3$ dimensional partition. 

Although we can not give a general formula of
$Y_D$ as in \eq{eqSUM}, we can directly confirm\footnote{
In MNS \cite{r:MNS}, it was proved by using totally different
argument.}
the corresponding equation,
\begin{equation}
\frac{1}{N}\sum_{|D|=N} Y_D=\sum_{d|N} \frac{1}{d^2},
\end{equation}
for $N=2,3,4$ cases by using symbolic computation.
Some of the explicit forms of $Y_D$ are given in the appendix.
The factor appear in various places, for example,
the computation of Witten index \cite{r:WI}, 
the coefficient of $R^4$ term
in the effective action of type IIB string 
theory \cite{r:RF} and so on.

The generating functional of $3$-dimensional partition
was once conjectured as,
\begin{equation}
 \sum_{n=0}^\infty \mu_3(n)q^n = \prod_{i=1}^\infty
  (1-q^i)^{-(i^2+i)/2}=1+q+4q^2+10q^3+26q^4+59q^5+141q^6+\cdots,
\end{equation}
where $\mu_3(n)$ should be
identified with the number of $3$-dimensional partition
of $n$.
If this formula were correct, $3$-dimensional partition can
be basically generated by creation operators with three indices
such as $\phi^\dagger_{ijk}$ with $i>0, j,k\geq 0$
(the degree is defined by $i+j+k$).
However, it is known that
for $n>5$, this formula disagree with the actual
number of partition \cite{r:And},
\begin{equation}
 \sum_{n=0}^\infty M_3(n)q^n
=1+q+4q^2+10q^3+26q^4+59q^5+142q^6+\cdots.
\end{equation}
The existence of oscillator representation
is natural from viewpoint in the next section.
However, there exit some additional 
structures which are at this moment beyond our reach.

\section{Jack polynomial in matrix theory}

It is tempting to imagine that there is the free field 
structure behind MNS calculation in $D=6$ case
since the contribution from each pole are parameterized
by Young table.
We would like to give a proposal.
A merit to give such a reinterpretation is
to illuminate the structure which is similar to the
matrix model\cite{r:old}.  It is well-known that old matrix model
has a description in terms of free bosons (or corresponding
free fermions).  This structure is intimately related to
the existence of the Virasoro (W) constraint
satisfied by the generating function,
\begin{equation}\label{eqgenfunc}
Z([t])\equiv <\exp(\sum_{n=1}^\infty t_n \mbox{Tr} \phi^n)>.
\end{equation}
It is interesting to investigate precise
form of the Virasoro structure in the new situation
since it will be intimately related to the structure
of the micro string \cite{r:DVV}. 

Although these two theories are commonly described by
free boson/fermion system in two dimensions,
there are some essential differences.

In old matrix model,
one may interpret the matrix size $N$ as the
eigenvalue of (non-relativistic) fermion number operator.
The free boson operator can be described
in terms of coefficients of the 
matrices $\mbox{Tr} \phi^n$ since it does not
change fermion sea level.

On the other hand, in the new theory,
the free boson operator $a_{-n}$ is defined as the creation of 
(degenerate) D-instantons
which changes the size of matrix by $n$.
Therefore we need to identify $N$ as eigenvalue of $L_0$. 
One may say that the difference comes from the fact that
the moduli space is parameterized by Grassman manifold
in the old model and by Hilbert scheme in the new model.

From this viewpoint, it does not seem to be natural
to consider the expectation value of the form
\eq{eqgenfunc}
as the generating function with $\phi$ be
the matrix for the longitudinal direction.
Indeed, in MNS integral, the generating functions
can be easily evaluated as,
\begin{equation}
Z_N([t])= \sum_{|D|=N} Y_D \exp(\sum_{\ell=1}^\infty t_\ell f_D^{(\ell)}),
\quad
f_D^{(\ell)}=\sum_{i=1}^N \phi(i,D)^\ell.
\end{equation}
(Here $\phi(i,D)$ is the value of $\phi$ in $i$'s box in the
Young diagram $D$.)
At this moment, since we can not identify $a_{-n}$ as 
$\sum_{i=1}^N \phi_i^n$, we will not pursue this line
in the following.


We would like find an operator formalism which 
naturally produces \eq{eqYD}.
As we argued, the residue
\eq{eqYD} have the coefficient which typically 
appear in the inner product
of Jack polynomial \cite{r:St}\cite{r:Mac}.
Let us give a brief summary of Calogero-Sutherland equation.
It is defined by motion of $N$ particles on a circle $0\leq q_i<L$.
Equation of motion is defined by the Hamiltonian,
\begin{equation}
  \widetilde{\cal H}_\beta=\frac{1}{2}\sum_{j=1}^N
  \biggl(\frac{1}{i}\frac{\partial}{\partial q_j}\biggr)^2
  +\frac{1}{2}\Bigl(\frac{\pi}{L}\Bigr)^2
  \sum_{i\neq j}\frac{\beta(\beta-1)}{\sin^2\frac{\pi}{L}(q_i-q_j)}.
\end{equation}
Let us introduce a new coordinate $x_j\equiv e^{2\pi iq_j/L}$
and rewrite Hamiltonian,
\begin{eqnarray}
  &&
  \widetilde{\Delta}(x)^{-\beta}\widetilde{\cal H}_\beta
  \widetilde{\Delta}(x)^{\beta}
  =2\Bigl(\frac{\pi}{L}\Bigr)^2 {\cal H}_\beta+E_0, \non
  &&
  {\cal H}_\beta\equiv
  \sum_{i=1}^N D_i^2
  +\beta\sum_{i<j}\frac{x_i+x_j}{x_i-x_j}(D_i-D_j),\\
  &&D_i\equiv x_i\frac{\partial}{\partial x_i}.
\label{e_H}
\end{eqnarray}
This Hamiltonian takes the space of symmetric polynomial of $x_i$
as the Hilbert space.
Eigenfunctions $\psi_\lambda(x)$ of ${\cal H}$
are called Jack symmetric
polynomials \cite{r:St}.
They are parameterized by Young diagrams and the
eigenvalue associated with the diagram
$\lambda=(\lambda_1,\cdots,\lambda_M)$
is given by,
\begin{equation}
\epsilon_{\lambda}=
  \sum_{i=1}^M\Bigl(\lambda_i^2+\beta(N+1-2i)\lambda_i\Bigr)
= \sum_{i=1}^{M'}\Bigl(-\beta\lambda_i^{\prime 2}
                       +(\beta N+2i-1)\lambda_i'\Bigr),
\end{equation}
where ${}^t\lambda=(\lambda_1',\cdots,\lambda_{M'}')$
is the transposed diagram of $\lambda$.

In order to relate this system with the matrix theory,
we need to introduce description in terms of 
free boson oscillators. Basic idea \cite{r:AMOS} was
to replace the power sum $\sum_{i=1}^N$ by boson oscillators
by using ket vector. More precisely, we use the following
mapping from free boson Fock space to space of symmetric
polynomials,
\begin{eqnarray}
  | f\rangle\mapsto
  f(x)
  &\!\!=\!\!&
  \langle C|f\rangle, \non
  C
  &\!\!=\!\!&
  \exp\left(\sqrt{\beta}\sum_{n>0}\frac{1}{n}a_np_n\right),\quad
  p_n=\sum_ix_i^n,
  \label{sfc}
\end{eqnarray}
The boson operators $a_n$ satisfies standard commutation
relation $[a_n,a_m]=n\delta_{n+m,0}$.
Under this mapping, the Hamiltonian \eq{e_H} is mapped to
\begin{eqnarray}
  H_{\beta}\langle 0 |  C& =&\langle 0 | C\hat{H}_{\beta},\non
 \hat{H}_{\beta}
  & = &
  \sqrt{\beta}\sum_{n,m>0}
  \Bigl(a_{-n-m}a_na_m
  +a_{-n}a_{-m}a_{n+m}\Bigr)
  +\sum_{n>0}a_{-n}a_n\Bigl((1-\beta)n+N\beta\Bigr)\non
&=& \sqrt{2\beta}\sum_{n>0}a_{-n}L_n
  +\sum_{n>0}a_{-n}a_n (N\beta+\beta-1-\sqrt{2\beta}a_0),
\label{e_HB}
\end{eqnarray}
where $L_n$ is Coulomb-gas representation of Virasoro generators
with the central charge $c=1-\frac{6(1-\beta)^2}{\beta}$.
As it may be easily seen from this form, 
oscillator representation of Jack polynomial
is closely related to the null vector of 
${\cal W}$-algebras \cite{r:AMOS}\cite{r:MY}.
In general, the explicit form of Jack polynomial
for arbitrary Young diagram is rather complicated object.
However, when Young diagram consists of just one row 
or one column, the corresponding Jack polynomial can 
be written rather easily. They are the expansion coefficients of
the screening current of $c<1$ CFT,
\begin{eqnarray}
 e^{-\frac{1}{\sqrt{\beta}}\phi(x)}|0\rangle & = &  
  \sum_{n=0}^\infty x^n |(n),\beta\rangle\non
 e^{\sqrt{\beta}\phi(x)}|0\rangle & = &  
  \sum_{n=0}^\infty x^n |(1^n),\beta\rangle.
\end{eqnarray}
For $\beta=1$ case,
Jack polynomial reduces to Schur polynomial.
The corresponding state in the Fock space
is the fermionic state defined through Maya diagram.

Let us proceed to investigate how Jack polynomial
appears in matrix theory residue formula \eq{eqYD}.
We make an identification of the deformation parameters with
the coupling of CS system.
$\beta=-E_2/E_1$.  
It will be a good idea to start from $\beta=1$ case
which is reducible to free fermion system.
Indeed \eq{eqYD} simplified drastically.
Only the diagrams that has non-vanishing contribution
are labeled by Young diagram of the form
$(n,1,\cdots,1)$, namely a single hook.
Furthermore they depends only on the number of boxes,
$Y_{D}=1/|D|^2$. For each $N$, we have $N$ types of such diagram
and the sum is just $1/N$.  (This is a much simpler way to
prove the observation in \cite{r:MNS}.)
Let us suppose that each contribution is 
represented by inner product associated with fermionic
state.  One may rewrite this identity as,
\begin{eqnarray}
 \frac{1}{N}&=&\frac{1}{N^2}\langle 0 | a_N a_{-N}|0\rangle\nonumber\\
 &=& \sum_{D} Y_D\label{e_YD}\\
 Y_D & = & |\langle 0| \frac{a_N}{N} |D,\beta=1\rangle|^2.\nonumber
\end{eqnarray}  
Here $|D,\beta=1\rangle$ is the fermionic state associated with diagram $D$.
For a single hook, they can be written simply,
$|(N-r,1^r),\beta=1\rangle = \psi^\dagger_{3/2-N+r}\psi_{-r+1/2}$

Corresponding to the fact that the free boson oscillator acts
on the cohomology class of the generating space of $\Hil{N}$,
one may write down the generating function of partition 
functions of $SU(N)$ matrix model as,
\begin{equation}
\sum_{N=0}^\infty N Z_N q^N = \langle 0 | \phi(q) \phi(1)|0\rangle.
\end{equation}
The formal expansion parameter $q$ is mapped to the coordinate
of the world sheet of the second quantized theory.
In this form the Virasoro structure is quite manifest.
Namely, the application of derivative on the left hand side
gives the insertion of Virasoro operator on the right hand side,
\begin{equation}
q^{n+1}\frac{\partial}{\partial q} \sum_{N=0}^\infty N Z_N q^N
= \langle 0 | [L_n,\phi(q)] \phi(1)|0\rangle.
\end{equation}

Another merit to write $Y_D$ as an inner product is
to make the independence on the deformation
parameter $\beta=-E_2/E_1$ manifest.  In \eq{e_YD},
one uses the decomposition of one in terms of free fermion
Hilbert space (or equivalently Schur polynomial).
In this language, the introduction of $\beta$ is simply to 
replace this bases to those defined by Jack polynomial
with a slight modification of ket vector.
\begin{eqnarray}
 \frac{1}{N}&=&\frac{1}{N}\langle 0 | a_N 
X_n|0\rangle\nonumber\\
 &=&  \sum_{|D|=N} Y_D\label{e_J}\\
 \sum_{n=0}^\infty q^n X_n|0\rangle
 & = & \frac{-1}{\sqrt{\beta}-\sqrt{1/\beta}}
e^{-(\sqrt{\beta}-\sqrt{1/\beta})\phi(q)}|0\rangle\\
 Y_D & = & \frac{\langle 0| \frac{a_N}{N} |D,\beta\rangle 
\langle D,\beta|X_n|0\rangle}{
\langle D,\beta|D,\beta \rangle}.\nonumber
\end{eqnarray}  
In this form, the incorporation of the normalization factor
of Jack polynomial in \eq{eqSUM} becomes quite natural.
Some non-triviality comes from matching the numerator.
For the Young diagrams of one row or one columns,
one may easily confirm this formula by using\footnote{
In this calculation in this section, 
we may drop the zero model of the
free boson $\phi$.},
\begin{eqnarray}
\langle 0 | e^{-\frac{1}{\sqrt{\beta}}\phi(t)}e^{-(\sqrt\beta
-\frac{1}{\sqrt\beta})\phi(t')}|0\rangle & = &
(1-tt')^{\frac{1}{\beta}-1}\non
\langle 0 | e^{\sqrt{\beta}\phi(t)}e^{-(\sqrt\beta
-\frac{1}{\sqrt\beta})\phi(t')}|0\rangle & = &
(1-tt')^{\beta-1}.
\end{eqnarray}

\section{Discussion}
Our discussion is still far from complete.
For example, although the appearance of Calogero-Sutherland
system is natural from MNS residue formula,
we can not derive it from the matrix integral itself.

One possible hint may come from the work of Lehn \cite{r:Lehn}
where he expressed the intersection with ``boundary'' 
(coinciding brane limit) of the
Hilbert scheme in terms of free boson operator. This operator
$\delta$ satisfies commutation relation with the oscillator as
\begin{equation}
[ \delta, a_n] = L_n.
\end{equation}
This commutation relation coincides with that of
Calogero-Sutherland Hamiltonian \eq{e_HB}.
In this connection, we have to also mention that
Nakajima \cite{r:Nakajima2} established the connection
between the Hilbert scheme of points on tangent space
of Riemannian surface with the Jack polynomial.

Another interesting problem is $D=10$ case.
As we illustrate it in appendix, the independence
on the deformation parameters are far from obvious.
Although the property of three dimensional partition
is rather exotic, 
we hope that the relation of the matrix equation,
\begin{equation}
[B_i,B_j]=\frac{1}{2}\epsilon_{ijkl}[B^\dagger_k,B^\dagger_l],
\end{equation}
with the self-dual equation in eight dimension\cite{r:GSD}
will give us some insight in the problem.

\vskip 5mm
\noindent{\bf Acknowledgement:}
We are obliged to T. Eguchi and M. Jinzenji for 
discussions and comments.

\section*{Appendix: Some Explicit calculations for D=10 case}

\noindent{\bf 2 boxes:}

\begin{enumerate}
\item Pole at $\phi_2=\phi_1+E_i$ ($i=1..4$). \newline
Residue$=\prod_{j(\neq i)}^4\frac{2E_i+E_j}{E_i-E_j}$.
\end{enumerate}
\begin{equation}
\sum_{i=1}^4 \prod_{j(\neq i)}^4\frac{2E_i+E_j}{E_i-E_j} =\frac{5}{2}.
\end{equation}

\vskip 2mm

\noindent {\bf 3 boxes:}
\begin{enumerate}
\item $\phi_3-\phi_2=\phi_2-\phi_1=E_i$.\newline
Residue$\equiv F^{(3)}_1[i]=\frac{1}{3}
\prod_{j(\neq i)}^4\frac{(2E_i+E_j)(3E_i+E_j)}{(E_i-E_j)(2E_i-E_j)}$.
\item $\phi_3-\phi_1=E_i$, $\phi_2-\phi_1=E_j$. We write its
      contribution as $F^{(3)}_2[i,j]$. For $i=1,j=2$, we have,
\begin{equation}
F^{(3)}_2[1,2]=\frac{(2E_1+E_2)(E_1+E_2)(E_1-2E_2-E_3)(E_1+E_3)
(2E_1-E_2+E_3)(E_2+E_3)}{(E_1-2E_2)(2E_1-E_2)(E_1-E_3)(E_2-E_3)
(E_1-E_4)(E_2-E_4)}
\end{equation}
\end{enumerate}
The summation:
\begin{equation}
\sum_{i=1}^4 F^{(3)}_1[i]+\sum_{j<i}F^{(3)}_2[i,j]=\frac{10}{3}.
\end{equation}


%
\end{document}